\documentclass[aps,prl,showpacs]{revtex4}
\usepackage{graphicx}
\usepackage{amsfonts,amssymb,amsmath}
\usepackage{bm}

\begin{document}

\title{Realizations of the $osp(2,1)$ Superalgebra and Related Physical
Systems}
\date{\today}
\author{Hayriye T\"{u}t\"{u}nc\"{u}ler}
\email{tutunculer@gantep.edu.tr}
\affiliation{Department of Physics, Faculty of Engineering 
University of Gaziantep,  27310 Gaziantep, Turkey}
\author{Ramazan Ko\c{c}}
\email{koc@gantep.edu.tr}
\affiliation{Department of Physics, Faculty of Engineering 
University of Gaziantep,  27310 Gaziantep, Turkey}

\begin{abstract}
Eigenvalues and eigenfunction of two-boson $2\times 2$ Hamiltonians in the
framework of the superalgebra $osp(2,1)$ are determined by presenting a
similarity transformation. The Hamiltonians include two bosons and one
fermion have been transformed in the form of the one variable differential
equations and the conditions for its solvability have been discussed. It is
observed that the Hamiltonians of the various physical systems can be
written in terms of the generators of the $osp(2,1)$ superalgebra and under
some certain conditions their eigenstates can exactly be obtained. In
particular, the procedure given here is useful in determining eigenstates of
the Jaynes-Cummings Hamiltonians.
\end{abstract}
\maketitle
\section{Introduction}

The Lie (super)algebras have played important role in the study of quantum
physics in particular they are associated with the symmetry properties of
physical systems and improve the understanding of physical structures
\cite{van, balan, radi, wes}. In the last decade a lot of effort, has been attracted
on the quasi-exactly solvable(QES) equations whose finite number of
eigenvector can be obtained by solving an algebraic equation
\cite{turb, shif, brihaye, ush, shaf}. Yet, even today, new contributions to this
problem are being made. They appear, however, not to have been fully
exploited in the analysis of QES equations. These systems have found
application in the different fields of the physics. Recently, great
attention is being paid to examine different quantum optical models
\cite{kara1, kara2, du, alv, deb} with Hamiltonians given by nonlinear functions of
the bosonic and/or fermionic operators since they enable to reveal new
effects and phenomena.

The natural step to relate the quantum optical systems and Lie
(super)algebras is to express the generators of the Lie (super)algebra in
terms of bosons and/or fermions. The Hamiltonians can be written as
combinations of the generators of a relevant symmetry group. Hence, one can
able to compute (a part of) the spectrum by performing a suitable
transformation of the generators.

One major symmetry group candidates for system of two differential equation
and $2\times 2$ matrix Hamiltonians is the supergroup $osp(2,1)$. It is well
known that unlike a purely bosonic algebra, the superalgebra admits different
Weyl inequivalent choices of simple root systems, which corresponds to
inequivalent Dynkin-diagrams. In the case of $osp(2,1)$ one has two choices
of simple roots which are unrelated by Weyl transformations: a system of
fermionic and bosonic simple roots, or a purely fermionic simple roots
\cite{ding}. Our aim, in this paper, is to construct bosonic and fermionic
realization of the $osp(2,1)$ algebra and to obtain solution of the some
quantum mechanical problems by performing a suitable transformation
operation for the bosons and fermions.

The paper is organized as follows: In section 2 the construction of the
boson-fermion realization of the $osp(2,1)$ algebra briefly reviewed and two
different realization have been given. Section 3 includes the general method
to transform the boson and fermion operators in the form of the QES. In
section 4 we present the solution of the Jaynes-Cummings Hamiltonian with
Kerr nonlinearity and modified-Jaynes-Cummings Hamiltonian as an application
of the method. The paper ends with a brief conclusion.

\section{Construction of the two-boson one fermion $osp(2,1)$ superalgebra}

A convenient way to construct a spectrum generating superalgebra for systems
with a finite number of bound states is by introducing a set of boson and
fermion operators. We introduce two boson operator, $a_{1}$ and $a_{2}$,
which obey the usual commutation relations%
\begin{equation}
	\left[ a_{1},a_{1}\right] =\left[ a_{1},a_{2}^{+}\right] =\left[
	a_{2},a_{1}^{+}\right] =0,\quad \left[ a_{1},a_{1}^{+}\right] =\left[
	a_{2},a_{2}^{+}\right] =1.  \label{1}
\end{equation}

The bilinear combinations $a_{1}^{+}a_{1}$, $a_{1}^{+}a_{2}$, $a_{2}^{+}a_{1}
$ and $a_{2}^{+}a_{2}$ generate the group $su(2)$ and $a_{1}^{+}a_{1}$, $%
a_{1}^{+}a_{2}^{+}$, $a_{2}a_{1}$ and $a_{2}^{+}a_{2}$ generate the group $%
su(1,1)$. Let us start by introducing three generators of $su(2)$,%
\begin{equation}
J_{+}=a_{1}^{+}a_{2},\quad J_{-}=a_{2}^{+}a_{1},\quad J_{0}=\frac{1}{2}%
\left( a_{1}^{+}a_{1}-a_{2}^{+}a_{2}\right) .  \label{2}
\end{equation}%
These are the Schwinger representation of $su(2)$ algebra. The fourth
generator is the total boson number operator%
\begin{equation}
N=a_{1}^{+}a_{1}+a_{2}^{+}a_{2}  \label{3}
\end{equation}%
which commutes with the $su(2)$ generators. The superalgebra $osp(2,1)$
might be constructed by extending $su(2)$ algebra with the fermionic
generators
\begin{subequations}
	\begin{eqnarray}
		V_{+} &=&f^{+}a_{2},V_{-}=-f^{+}a_{1},W_{+}=fa_{1}^{+},W_{-}=fa_{2}^{+}
		\label{4a} \\
		V_{+} &=&fa_{2},V_{-}=-fa_{1},W_{+}=f^{+}a_{1}^{+},W_{-}=f^{+}a_{2}^{+},
		\label{4b}
	\end{eqnarray}%
	where $f^{+}$ and $f$ are fermions and they satisfy the anticommutation
	relation
\end{subequations}
\begin{equation}
	\left\{ f,f^{+}\right\} =1.  \label{5}
\end{equation}%
The superalgebra $osp(2,1)$ can be constructed with the generators (\ref{2})
and (\ref{4a}) or (\ref{4b}). It is easily seen that the generators given in
(\ref{4a}) and (\ref{4b}) can be mapped on to each others by a change of the
fermionic creation and annihilation operators. As discussed in
\cite{chen1, chen2}, the generators of the $osp(2,1)$ algebra are written as follows:
\begin{equation}
	\left\{ J_{\pm },J_{0},J\in osp(2,1)_{\bar{0}}\quad
		{\vert}%
	\quad V_{\pm },W_{\pm }\in osp(2,1)_{\bar{1}}\right\},   \label{6}
\end{equation}%
where $J$ is the total number operator of the system and is given by
\begin{subequations}
	\begin{eqnarray}
		J &=&\frac{1}{2}N+f^{+}f  \label{7a} \\
		J &=&\frac{1}{2}N+ff^{+}  \label{7b}
	\end{eqnarray}
	for the generators (\ref{2}, \ref{4a}) and (\ref{2}, \ref{4b}), respectively.
	The generators of the $osp(2,1)$ superalgebra satisfy the following
	commutation and anticommutation relations:
\end{subequations}
\begin{eqnarray}
	\left[ J_{+},J_{-}\right]  &=&2J_{0},\quad \quad \left[ J_{0},J_{\pm }\right]
	=\pm J_{\pm },\quad \quad \quad \left[ J,J_{\pm ,0}\right] =0  \notag \\
	\left[ J_{0},V_{\pm }\right]  &=&\pm \frac{1}{2}V_{\pm },\quad \left[
	J_{0},W_{\pm }\right] =\pm \frac{1}{2}W_{\pm },\quad \left[ J,V_{\pm }\right]
	=\frac{1}{2}V_{\pm }  \notag \\
	\left[ J_{\pm },V_{\mp }\right]  &=&V_{\pm },\quad \quad \left[ J_{\pm
	},W_{\mp }\right] =W_{\pm },\quad \quad \left[ J,W_{\pm }\right] =-\frac{1}{2%
	}W_{\pm }  \notag \\
	\left[ J_{\pm },V_{\pm }\right]  &=&0,\quad \quad \quad \left[ J_{\pm
	},W_{\pm }\right] =0  \label{8} \\
	\left\{ V_{\pm },W_{\pm }\right\}  &=&\pm Q_{\pm },\quad \left\{ V_{\pm
	},W_{\mp }\right\} =-J_{0}\pm J  \notag \\
	\left\{ V_{\pm },V_{\pm }\right\}  &=&\left\{ V_{\pm },V_{\mp }\right\}
	=\left\{ W_{\pm },W_{\pm }\right\} =\left\{ W_{\pm },W_{\mp }\right\} =0.
	\notag
\end{eqnarray}%
It is well known that the $osp(2,1)$ algebra has been constructed by extending
purely bosonic $su(2)$ algebra with fermionic generators. Meanwhile, we
mention here, one can construct $osp(2,2)$ algebra by extending $su(1,1)$
algebra.

In general, the spectrum of the physical system can be calculated in a
closed form when the Hamiltonian of the system can be written in terms of
number operator and diagonal operator $J_{0}$, it can be diagonalized within
the representation $[N]$. The abstract boson and/or fermion algebra can be
associated with the exactly solvable Schr\"{o}dinger equations by using the
differential operator realizations of boson operators. This connection opens
the way to an algebraic treatment of a large class of potentials of
practical interest \cite{gursey}.

The Hamiltonian of the some physical systems can be constructed up to
quadratic combination of the generators of the $osp(2,1)$ algebra. It will
be shown that the associated Hamiltonian includes various physical system
Hamiltonians. In order to relate the $osp(2,1)$ algebraic structure with the
two dimensional $2\times 2$ matrix Hamiltonians one can use standard matrix
representations of the fermions:%
\begin{eqnarray}
&&f=\sigma _{-}=\left(
\begin{array}{cc}
0 & 0 \\
1 & 0%
\end{array}%
\right) ,\quad f^{+}=\sigma _{+}=\left(
\begin{array}{cc}
0 & 1 \\
0 & 0%
\end{array}%
\right) ,\quad   \label{9} \\
&&ff^{+}-f^{+}f=\sigma _{0}=\left(
\begin{array}{cc}
1 & 0 \\
0 & -1%
\end{array}%
\right) .  \notag
\end{eqnarray}%
The bosons can be realized in the Bargmann-Fock space. In order to obtain a
(quasi)exactly solvable Hamiltonian, in the next section, we prepare a
suitable transformation procedure.

\section{Transformation of the operators}

As discussed in \cite{gursey}, the Schwinger representation of $su(2)$ and $%
su(1,1)$ algebra with two boson operator can be mapped onto the generalized
Gelfand-Dyson representation by a similarity transformation. The connection
between the two representation can be obtained by introducing the operators
\begin{equation}
\Gamma _{1}=(a_{2})^{a_{1}^{+}a_{1}}\text{ \ or \ }\Gamma
_{2}=(a_{2}^{+})^{a_{1}^{+}a_{1}}.  \label{e1}
\end{equation}%
Since $a_{1}$ and $a_{2}$ commute, $\Gamma _{1}$ and $\Gamma _{2}$ are well
defined and their actions on the state are given by
\begin{subequations}
\begin{eqnarray}
\Gamma _{1}\left| n_{1},n_{2}\right\rangle  &=&(a_{2})^{n_{1}}\left|
n_{1},n_{2}\right\rangle =\sqrt{\frac{n_{2}!}{(n_{2}-n_{1})!}}\left|
n_{1},n_{2}-n_{1}\right\rangle  \\
\Gamma _{2}\left| n_{1},n_{2}\right\rangle  &=&(a_{2}^{+})^{n_{1}}\left|
n_{1},n_{2}\right\rangle =\sqrt{\frac{n_{2}!}{(n_{2}+n_{1}+1)!}}\left|
n_{1},n_{2}+n_{1}\right\rangle .
\end{eqnarray}%
Our task is now to develop a transformation procedure to obtain generalized
Gelfand-Dyson \textit{like} representation of the $osp(2,1)$ superalgebra.
This can be done by introducing the following similarity transformation
induced by the metric:
\end{subequations}
\begin{equation}
	`S=(a_{2}^{+})^{a_{1}^{+}a_{1}+\alpha \sigma _{+}\sigma _{-}},  \label{10}
\end{equation}%
where $\alpha $ takes the values $\pm 1$ . Since $a_{1}$, and $a_{2}$
commute and $\sigma _{\pm ,0}$ also commute with the bosonic operators, the
transformation of $a_{1}$ and $a_{1}^{+}$ under $S$ can be obtained by
writing $a_{2}^{+}=e^{b}$, with $[a_{1},b]=[a_{1}^{+},b]=0$,%
\begin{eqnarray}
Sa_{1}S^{-1} &=&a_{1}(a_{2}^{+})^{-1}  \notag \\
Sa_{1}^{+}S^{-1} &=&a_{1}^{+}a_{2}^{+};  \label{11}
\end{eqnarray}%
the transformations of $a_{2}$ and $a_{2}^{+}$ are
\begin{eqnarray}
	Sa_{2}S^{-1} &=&a_{2}-n(a_{2}^{+})^{-1}  \notag \\
	Sa_{2}^{+}S^{-1} &=&a_{2}^{+},  \label{12}
\end{eqnarray}%
and the transformations of the $\sigma _{\pm }$ are given by%
\begin{equation}
S\sigma _{\pm }S^{-1}=\sigma _{\pm }(a_{2}^{+})^{\pm \alpha }  \label{13}
\end{equation}%
where $n$ is given by%
\begin{equation}
n=a_{1}^{+}a_{1}+\alpha \sigma _{+}\sigma _{-}\,.  \label{14}
\end{equation}

The transformations of the bosons and fermions (\ref{11}) through (\ref{14})
play a key role in the construction of QES one-variable $2\times 2$ matrix
Hamiltonians. For two different values of $\alpha =\pm 1$, the two
component polynomial spinors form a basis function for the generators of the
$osp(2,1)$ algebra. Consequently, we obtain two classes of Hamiltonians which
can be solved (quasi)exactly under some certain conditions.

\subsection{Case: $S=(a_{2}^{+})^{a_{1}^{+}a_{1}+\protect\alpha \protect%
\sigma _{+}\protect\sigma _{-}};$ $\protect\alpha =1$}

In the case of $\alpha =1$, under the transformations (\ref{11}--\ref{14}),
the generators (\ref{2}), (\ref{4a}) and (\ref{7a}) of the $osp(2,1)$
algebra take the form%
\begin{eqnarray}
J_{+}^{\prime }
&=&SJ_{+}S^{-1}=-a_{1}^{+}a_{1}^{+}a_{1}+a_{1}^{+}(a_{2}^{+}a_{2}-\sigma
_{+}\sigma _{-})  \notag \\
J_{-}^{\prime } &=&SJ_{-}S^{-1}=a_{1}  \notag \\
J_{0}^{\prime } &=&SJ_{0}S^{-1}=\frac{1}{2}\left(
2a_{1}^{+}a_{1}-a_{2}^{+}a_{2}+\sigma _{+}\sigma _{-}\right)  \notag \\
J^{\prime } &=&SJS^{-1}=\frac{1}{2}\left( a_{2}^{+}a_{2}+\sigma _{+}\sigma
_{-}\right)  \notag \\
V_{+}^{\prime } &=&SV_{+}S^{-1}=\sigma
_{+}(a_{2}^{+}a_{2}-a_{1}^{+}a_{1}-\sigma _{+}\sigma _{-})  \label{15} \\
V_{-}^{\prime } &=&SV_{-}S^{-1}=-\sigma _{+}a_{1}  \notag \\
W_{+}^{\prime } &=&SW_{+}S^{-1}=\sigma _{-}a_{1}^{+}  \notag \\
W_{-}^{\prime } &=&SW_{-}S^{-1}=\sigma _{-}\,.  \notag
\end{eqnarray}

The difference between the representations of $osp(2,1)$ given in section 1
and primed representations (\ref{15}) is that, while in the first the total
number of $a_{1}$, $a_{2}$ bosons and $f$, $f^{+}$ fermions characterize the
the system, in the later it is only the number of $a_{2}$ bosons that
characterize the system. When we characterize the algebra by a fixed number $%
a_{2}^{+}a_{2}=j$ in the primed representation, the generators can be
expressed in terms of one-boson operator $a_{1}$ and yield the following
realization:%
\begin{eqnarray}
	J_{+}^{\prime } &=&SJ_{+}S^{-1}=-a_{1}^{+}a_{1}^{+}a_{1}+a_{1}^{+}(j-\sigma
	_{+}\sigma _{-})  \notag \\
	J_{-}^{\prime } &=&SJ_{-}S^{-1}=a_{1}  \notag \\
	J_{0}^{\prime } &=&SJ_{0}S^{-1}=\frac{1}{2}\left( 2a_{1}^{+}a_{1}-j+\sigma
	_{+}\sigma _{-}\right)  \notag \\
	J^{\prime } &=&SJS^{-1}=\frac{1}{2}\left( j+\sigma _{+}\sigma _{-}\right)
	\notag \\
	V_{+}^{\prime } &=&SV_{+}S^{-1}=-\sigma _{+}(a_{1}^{+}a_{1}-j)  \label{16} \\
	V_{-}^{\prime } &=&SV_{-}S^{-1}=-\sigma _{+}a_{1}  \notag \\
	W_{+}^{\prime } &=&SW_{+}S^{-1}=\sigma _{-}a_{1}^{+}  \notag \\
	W_{-}^{\prime } &=&SW_{-}S^{-1}=\sigma _{-}\,.  \notag
\end{eqnarray}%
These generators play an important role in the quasi-exact solution of the
matrix Schr\"{o}dinger equation. The generators of the $osp(2,1)$ algebra
can be expressed as differential equation in the Bargmann-Fock space by
defining the bosonic operators:
\begin{equation}
	a_{1}=\frac{d}{dx},\quad a_{1}^{+}=x.  \label{17}
\end{equation}

The two component polynomials of degree $j$ and $j+1$ form a basis function
for the generators of the $osp(2,1)$ algebra in the Bargmann-Fock space:
\begin{equation}
	P_{n+1,n}(x)=\left(
	\begin{array}{c}
		x^{0},x^{1},\cdots ,x^{n+1} \\
		x^{0},x^{1},\cdots ,x^{n}%
	\end{array}%
	\right).  \label{18}
\end{equation}

The general QES operator can be obtained by linear and bilinear combinations
of the generators of the $osp(2,1)$ superalgebra. Action of the QES operator
on the basis function (\ref{18}) gives us a recurrence relation; therefore,
the wavefunction is itself the generating function of the energy
polynomials. The eigenvalues are then produced by the roots of such
polynomials. Before illustrating this application of the procedure given
here on the physical examples, let us construct another representations of
the $osp(2,1)$ superalgebra.

\subsection{Case: $S=(a_{2}^{+})^{a_{1}^{+}a_{1}+\protect\alpha \protect%
			\sigma _{+}\protect\sigma _{-}};$ $\protect\alpha =-1$}

Using the same similarity transformation procedure given in section 3, we
obtain the second class of the $osp(2,1)$ superalgebra. In this
case, the generators $J_{\pm ,0}$ remain the same as in (\ref{16}), while the
generators $V_{\pm },W_{\pm }$ and $J$ given in (\ref{4b}) and (\ref{7b}),
respectively, take the form
\begin{eqnarray}
	J^{\prime } &=&SJS^{-1}=\frac{1}{2}\left( j+1+\sigma _{-}\sigma _{+}\right)
	\notag \\
	V_{+}^{\prime } &=&SV_{+}S^{-1}=-\sigma _{-}(a_{1}^{+}a_{1}-j-1)  \notag \\
	V_{-}^{\prime } &=&SV_{-}S^{-1}=-\sigma _{-}a_{1}  \label{19} \\
	W_{+}^{\prime } &=&SW_{+}S^{-1}=\sigma _{+}a_{1}^{+}  \notag \\
	W_{-}^{\prime } &=&SW_{-}S^{-1}=\sigma _{+}\,.  \notag
\end{eqnarray}

The basis function of this structure takes the form

$\qquad $%
\begin{equation}
	P_{n,n+1}(x)=\left(
	\begin{array}{c}
		x^{0},x^{1},\cdots ,x^{n} \\
		x^{0},x^{1},\cdots ,x^{n+1}%
	\end{array}%
	\right).  \label{20}
\end{equation}

The other transformation can be done by introducing the following similarity
transformation induced by the metric
\begin{equation}
	T=(a_{2})^{-a_{1}^{+}a_{1}+\eta \sigma _{+}\sigma _{-}},  \label{21}
\end{equation}%
where $\eta $ takes the values $\pm 1$. By using the similar arguments
given in the previous section one can easily obtain the following
transformations:%
\begin{eqnarray}
	Ta_{1}T^{-1} &=&a_{1}a_{2}^{+}  \notag \\
	Ta_{1}^{+}T^{-1} &=&a_{1}^{+}(a_{2}^{+})^{-1}  \notag \\
	Ta_{2}T^{-1} &=&a_{2}  \label{22} \\
	Ta_{2}^{+}T^{-1} &=&a_{2}^{+}+n(a_{2}^{+})^{-1}  \notag \\
	T\sigma _{\pm }T^{-1} &=&\sigma _{\pm }(a_{2}^{+})^{\pm \alpha }.  \notag
\end{eqnarray}

With this transformation we can construct two more different realizations
of the $osp(2,1)$ algebra for $\eta =\pm 1$.

\subsection{Case: $T=(a_{2})^{-a_{1}^{+}a_{1}+\protect\eta \protect\sigma %
_{+}\protect\sigma _{-}};\quad \protect\eta =1$}

In the case of $\eta =1$ , under the transformations (\ref{22}) the
generators (\ref{2}), (\ref{4b}) and (\ref{7b}) of the $osp(2,1)$ algebra
take the form%
\begin{eqnarray}
J_{+}^{\prime } &=&TJ_{+}T^{-1}=a_{1}^{+}  \notag \\
J_{-}^{\prime } &=&TJ_{-}T^{-1}=a_{1}^{+}a_{1}a_{1}+(a_{2}^{+}a_{2}+\sigma
_{+}\sigma _{-})  \notag \\
J_{0}^{\prime } &=&TJ_{0}T^{-1}=\frac{1}{2}\left(
2a_{1}^{+}a_{1}-a_{2}^{+}a_{2}-\sigma _{+}\sigma _{-}\right)  \notag \\
J^{\prime } &=&TJT^{-1}=\frac{1}{2}\left( a_{2}^{+}a_{2}+1+\sigma _{-}\sigma
_{+}\right)  \notag \\
V_{+}^{\prime } &=&TV_{+}T^{-1}=\sigma _{-}  \label{23} \\
V_{-}^{\prime } &=&TV_{-}T^{-1}=-\sigma _{-}a_{1}  \notag \\
W_{+}^{\prime } &=&TW_{+}T^{-1}=\sigma _{+}a_{1}^{+}  \notag \\
W_{-}^{\prime } &=&TW_{-}T^{-1}=\sigma
_{+}(a_{2}^{+}a_{2}-a_{1}^{+}a_{1}+1+\sigma _{+}\sigma _{-}).  \notag
\end{eqnarray}

This realization can also be characterized by $a_{2}^{+}a_{2}=j$. The basis
function of the realization is given by
\begin{equation}
	P_{n+1,n}(x)=\left(
	\begin{array}{c}
		x^{0},x^{1},\cdots ,x^{n+1} \\
		x^{0},x^{1},\cdots ,x^{n}%
	\end{array}%
	\right),  \label{24}
\end{equation}
in the Bargmann-Fock space.

\subsection{Case: $T=(a_{2})^{-a_{1}^{+}a_{1}+\protect\eta \protect\sigma %
_{+}\protect\sigma _{-}};\quad \protect\eta =-1$}

The generators (\ref{2}) of the $osp(2,1)$ superalgebra take the same form
as in (\ref{23}) and the remaining generators can be expressed as:%
\begin{eqnarray}
	J^{\prime } &=&TJT^{-1}=\frac{1}{2}\left( a_{2}^{+}a_{2}-\sigma _{+}\sigma
	_{-}\right)   \notag \\
	V_{+}^{\prime } &=&TV_{+}T^{-1}=\sigma _{+}  \label{25} \\
	V_{-}^{\prime } &=&TV_{-}T^{-1}=-\sigma _{+}a_{1}  \notag \\
	W_{+}^{\prime } &=&TW_{+}T^{-1}=\sigma _{-}a_{1}^{+}  \notag \\
	W_{-}^{\prime } &=&TW_{-}T^{-1}=\sigma
	_{-}(a_{2}^{+}a_{2}-a_{1}^{+}a_{1}+1-\sigma _{+}\sigma _{-}).  \notag
\end{eqnarray}%
The basis function of this structure is given by%
\begin{equation}
	P_{n,n-1}(x)=\left(
	\begin{array}{c}
		x^{0},x^{1},\cdots ,x^{n} \\
		x^{0},x^{1},\cdots ,x^{n-1}%
	\end{array}%
	\right).   \label{25a}
\end{equation}%
Consequently we have obtained four classes generators for the $osp(2,1)$
superalgebra by using two transformation operators $S$ and $T$. These
generators can be expressed as one-variable $2\times 2$ matrix differential
operators useful in the study of (quasi) exactly solvable systems.

\section{Application}

This section includes solution of the some physical Hamiltonians by using
the procedure given in the previous sections of this article. In particular,
our approach is useful for the study of nonlinear optical systems.

\subsection{Jaynes-Cummings Hamiltonian with the Kerr nonlinearity}

The effective Hamiltonian, which represents the Jaynes-Cummings model with
Kerr nonlinearity, has been expressed as \cite{buzek}
\begin{equation}
	H=\omega a^{+}a+\frac{1}{2}\omega _{0}\sigma _{0}+\kappa (a^{+}\sigma
	_{-}+a\sigma _{+})+\lambda a^{+}aa^{+}a,  \label{26}
\end{equation}%
where $a$ and $a^{+}$ are annihilation and creation operators of the
radiation mode which, with a frequency $\sigma _{\pm ,0}$, are the standard
Pauli matrices for the atom, and has a frequency of transition $\omega _{0}
$; $\kappa $, $\lambda $ are the coupling constant of the field and atom and
the coupling constant of the field and Kerr medium, respectively. The eigenvalue
equation can be written as
\begin{equation}
	H\psi =E\psi .  \label{27}
\end{equation}%
Our task is now to express the Hamiltonian (\ref{26}) in terms of the
generators of the $osp(2,1)$ algebra. In terms of the generators given in
(\ref{2}), (\ref{4a}) and (\ref{7a}) and number operator $N$, the Hamiltonian can be
written as
\begin{equation}
	H^{\prime }=\omega (2J_{0}+N)+\frac{\omega _{0}}{2}\left( J-N-J_{0}-1\right)
	+\lambda (2J_{0}+N)^{2}+\kappa (W_{+}-V_{-}).  \label{28}
\end{equation}%
Note that the bosonic operators $a_{1}=a$ and $a_{1}^{+}=a^{+}.$ By
considering the transformation procedures given in (\ref{16}) one can obtain
the single variable differential equation, which in the Bargmann-Fock space reads:
\begin{equation}
	H^{\prime }=(\omega +\omega _{0})(2x\frac{d}{dx}+1-j)+(\omega -\omega
	_{0})\sigma _{0}+\lambda (2x\frac{d}{dx}+1+\sigma _{0}-j)^{2}+\kappa
	(x\sigma _{-}+\frac{d}{dx}\sigma _{+}).  \label{29}
\end{equation}%
The action of the Hamiltonian on the two component spinor%
\begin{equation}
	P_{n,m}(x)=\left(
	\begin{array}{c}
		u_{n}(x) \\
		v_{m}(x)%
	\end{array}%
	\right)   \label{30}
\end{equation}%
gives us the following recurrence relation:
\begin{subequations}
	\begin{eqnarray}
		\left( 2\omega -(\omega +\omega _{0})(j-2n)+\lambda (j-2(n+1))^{2}-E\right)
		u_{n}(E)+\kappa mv_{m-1}(E) &=&0  \label{31a} \\
		-\left( E+(j-2m)(\omega -\lambda (j-m)+(j-2m-2)\omega _{0}\right)
		v_{m}(E)+\kappa u_{n+1} &=&0.  \label{31b}
	\end{eqnarray}%
	The recurrence relation implies that the wavefunction is itself the
	generating function of the energy polynomials. The eigenvalues are then
	given by the roots of such polynomials. If $E_{j}$ is a root of the
	polynomial, the wavefunction is truncated at $n+1=m=j$ and belongs to the
	spectrum of the Hamiltonian $H$. Considering the initial condition $%
	v_{j+1}=u_{j+1}=0$, we have obtained first few eigenvalues of the
	Hamiltonian $H$:
\end{subequations}
\begin{equation*}
	\left(
	\begin{array}{cc}
		9\lambda +3\omega +\omega _{0}-E & 0 \\
		0 & \lambda -\omega +\omega _{0}-E%
	\end{array}%
	\right) \left(
	\begin{array}{c}
		u_{1} \\
		v_{0}%
	\end{array}%
	\right) = 0.
\end{equation*}
Therefore we can obtain
\begin{equation*}
	E=\{9\lambda +3\omega +\omega _{0},\quad \lambda -\omega +\omega _{0}\}
\end{equation*}%
for $j=1.$ Similarly for $j=2$, the eigenvalues of the Hamiltonian are given
by%
\begin{equation}
	E=\{2\omega _{0},\quad 2(\omega +2\lambda ),\quad \omega +10\lambda +\omega
	_{0}\pm \sqrt{\kappa ^{2}+(3\omega +6\lambda +\omega _{0})^{2}}.
\end{equation}%
Analytical solutions of the recurrence relations (\ref{31a}) and (\ref{31b})
are available only for the first few values of $\ j,$ for large $j$ the
solutions become numerical.

\subsection{Modified Jaynes-Cummings Hamiltonian}

The modified Jaynes-Cummings Hamiltonian has been constructed to investigate
single two level atom placed in the common domain of two cavities
interacting with two quantized modes. It is given by \cite{bo}
\begin{equation}
H=\omega (a_{1}^{+}a_{1}+a_{2}^{+}a_{2})+\frac{\omega _{0}}{2}\sigma
_{0}+\lambda _{1}(a_{1}\sigma _{+}+a_{1}^{+}\sigma _{-})+\lambda
_{2}(a_{2}\sigma _{+}+a_{2}^{+}\sigma _{-}).  \label{32}
\end{equation}%
Using the same procedure as in section 4.1, we can obtain the the
transformed form of the Hamiltonian (\ref{32}) in terms of the generators $%
osp(2,1)$ algebra, given in (\ref{2}), (\ref{4a}) and (\ref{7a}) and with number operator $%
N$; then the Hamiltonian can be written as

\begin{equation}
H=\omega N+\frac{\omega _{0}}{2}\left( J-1-\frac{N}{2}\right) +\lambda
_{1}(W_{+}-V_{-})+\lambda _{2}(W_{-}+V_{+}).  \label{33}
\end{equation}

The eigenvalues of this Hamiltonian can be obtained by using the
transformation procedure given (\ref{15}) or (\ref{25}). The transformed
Hamiltonian can be expressed as one variable differential operator:%
\begin{equation}
H^{\prime }=\omega (j-1-\sigma _{0})+\frac{\omega _{0}}{2}\sigma
_{0}+\lambda _{1}\left( x\sigma _{-}+\frac{d}{dx}\sigma _{+}\right) +\lambda
_{2}\left( \sigma _{\_}-(x\frac{d}{dx}-j)\sigma _{+}\right) .  \label{34}
\end{equation}%
Therefore two component one variable spinor (\ref{30}) from a basis function
for the Hamiltonian (\ref{34})%
\begin{eqnarray*}
	\phi _{1}(x) &=&C_{1}(\lambda _{2}+\lambda _{1}x)^{n}(\lambda _{1}-\lambda
	_{2}x)^{j-n} \\
	\phi _{2}(x) &=&C_{2}(\lambda _{2}+\lambda _{1}x)^{n-1}(\lambda _{1}-\lambda
	_{2}x)^{j-n}\,,
\end{eqnarray*}%
where $n=0,1,\cdots ,j,$ and eigenvalues of the Hamiltonian are given by%
\begin{equation*}
	E=\omega (j-1)\pm \sqrt{(\omega _{0}-2\omega )^{2}+4n(\lambda
	_{1}^{2}+\lambda _{2}^{2})}\,.
\end{equation*}%
Consequently we have obtained the exact result for the eigenvalues of the
modified Jaynes-Cummings Hamiltonian. The procedure given here can be
applied to obtain eigenfunction and eigenvalues of various physical
Hamiltonians. In this section we have discussed solution of the some simple
physical Hamiltonians by using the transformation of the $osp(2,1)$
superalgebra, without further details.

\section{Conclusion}

The basic features of our approach is to construct $osp(2,1)$ invariant
subspaces. We consider systems whose Hamiltonian can be expressed in terms
of two bosons and one fermions. Furthermore, we have presented two different
boson-fermion realization of $osp(2,1)$ algebra. The corresponding
realizations have been transformed in the form of one dimensional
differential equations. Meanwhile, we have shown that our procedure is
appropriate to obtain eigenvalues and eigenfunctions of various systems. In
particular, we have constructed solutions to the Jaynes-Cummings and Modified Jaynes-Cummings Hamiltonians.

The suggested approach can be generalized in various directions. Invariant
subspaces of the multi-boson and multi-fermion systems can be obtained by
extending the method given in this paper. In particular, the subspace of the $%
osp(2,2)$ superalgebra can easily be constructed.


\begin{thebibliography}{99}


\bibitem{van} Van der Jeugt J (1987) J. Math. Phys. 28 758
\bibitem{balan} Balantekin A B (1984) J. Math. Phys. 25 2028
\bibitem{radi} Radicliffe J M (1971) J. Phys. A: Math. Gen. 4 313
\bibitem{wes} Wess J and Zumino B (1974) Nucl. Phys. B78 1
\bibitem{turb} Turbiner A V (1988) Comm. Math. Phys. 119 467
\bibitem{shif} Shifman M A Trubiner A V (1989) Comm. Math. Phys. 120 347
\bibitem{brihaye} Brihaye Y, Kosinski P (1994) J. Math. Phys. 35 3089
\bibitem{ush} Ushveridze A G (1993) Quasi-exact solvability in quantum
mechanics, IOP publishing, Bristol and Philadelphia

\bibitem{shaf} Shafiekhani A and Khorrami M (1997) Mod. Phys. Lett. A12, 22
\bibitem{kara1} Karassiov V P (1998) Phys. Lett. A 238 19
\bibitem{kara2} Karassiov V P and Klimov A B (1994) Phys. Lett. A 189 43
\bibitem{du} Du Si-De, Gong Shang-qing, Xu Zhi-zhan and Gong Chang-de (1997)
Quantum Semiclass. Opt. 9 941

\bibitem{alv} Alvarez G, Finkel F, Gonzalez-Lopez A and Rodriguez M\ A
(2002) \ J. Phys. A: Math. Gen. 35, 8705

\bibitem{deb} Debergh N and Stancu Fl. (2001) J.Phys.A: Math. Gen 34 18

\bibitem{ding} Ding X M, Gould M D, Mewton C J, Zhang Y Z (2002)
hep-th/0211235

\bibitem{chen1} Chen Yong-Qing (2000) J. Phys. A: Math. Gen. 33 8071; (2001)
Int. J. Theor. Phys. 40 1113; (2000) Int. J. Theor. Phys. 39 2523

\bibitem{chen2} Chen Yong-Qing, Xiao-Hui Liu and Xing-Chang Song (1994)
Commun. Theor. Phys. 22 123

\bibitem{gursey} Alhassid Y, G\"{u}rsey F and Iachello F (1983) Annals of
Physics 148 346

\bibitem{buzek} Buzek V and Jex I 1990 Opt. Commun. 78 425

\bibitem{bo} Jing-Bo Xu and Xu-Bo Zou (2001) Chin. Phys. Lett. 18 51
	
\end{thebibliography}
\end{document}